\documentclass[11pt,twoside]{article}

\usepackage{asp2006_hotcool}
\usepackage{epsf}
\usepackage{epsfig}
\usepackage{lscape}
\usepackage{natbib}

\begin{document}
\title{Chromospheres and Winds of Red Supergiants:\\ An Empirical Look at 
       Outer Atmospheric Structure}   
\author{Philip D.\ Bennett\altaffilmark{1}$^,$\altaffilmark{2}} 
\altaffiltext{1}{Eureka Scientific, Inc., 2452 Delmer Street, Suite 100,
       Oakland, CA 94602-3017, USA}
\altaffiltext{2}{Saint Mary's University, Department of Astronomy and Physics,
       Halifax, NS ~B3H 3C3, Canada}    

\begin{abstract} 
Stars between about 4 and 25 solar masses spend a significant fraction of 
their post-main sequence lifetime as red supergiants (RSGs) and lose 
material via stellar winds during this period. For RSGs more massive than 
10 solar masses, this mass loss becomes of evolutionary significance, and 
probably determines the upper mass limit of RSGs in the Hertzsprung-Russell
diagram.
Despite decades of observations, the driving mechanism responsible for 
mass loss in RSGs remains unknown. Mainly this is because the optical 
spectrum accessible from the ground provides almost no useful wind 
diagnostics, and what information is obtained is spatially averaged over 
the entire wind volume. However, within the last decade, {\em Hubble Space 
Telescope (HST)} observations of many useful ultraviolet wind diagnostics 
have been obtained at a high signal-to-noise ratio and spectral resolution. In particular, 
RSGs in eclipsing binaries can provide spatially resolved observations 
of stellar chromospheres and winds. 
I review possible RSG wind acceleration mechanisms, discuss
some observational constraints, and present some empirical models 
of RSG chromospheres and winds.
\end{abstract}


\section{Stellar Winds in the H-R Diagram}

\begin{figure}[htb]
\plotfiddle{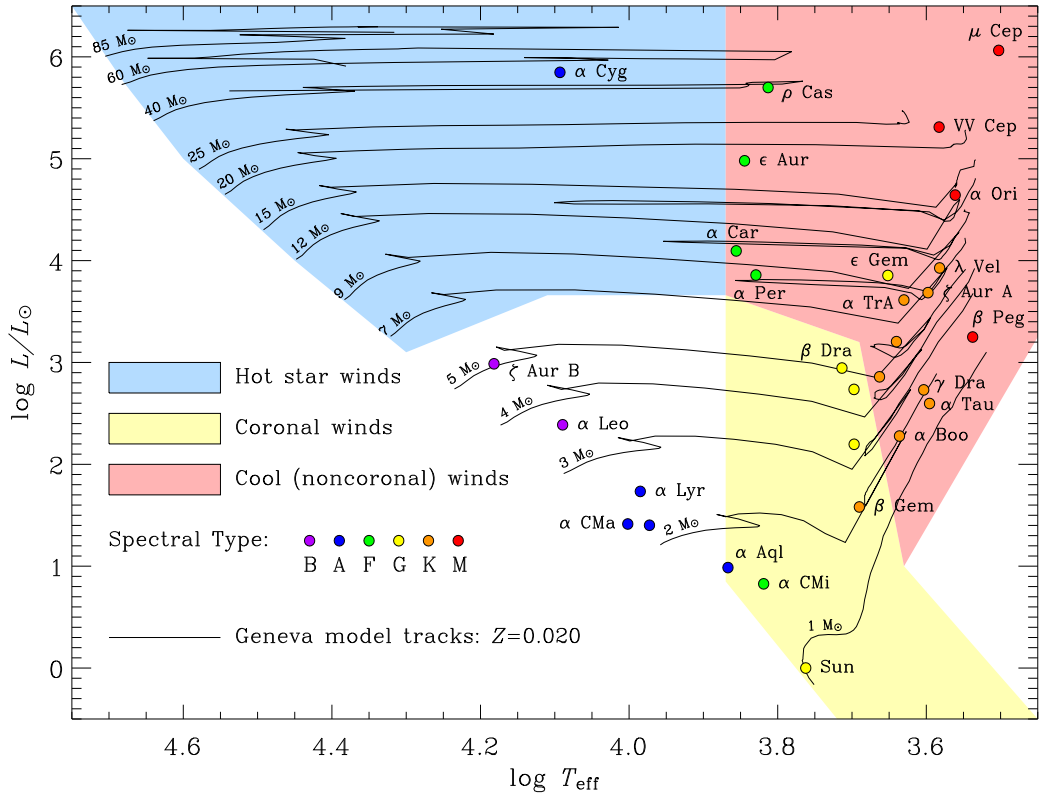}{3.9in}{90}{62}{64}{240}{-17}
\caption{Stellar Winds in the Hertzsprung-Russell Diagram.}
\label{fig:hr_diag}
\end{figure}

Stellar winds are a ubiquitous phenomenon across much of the 
Hertzsprung-Russell diagram (HRD). These winds can be
broadly grouped into the three main categories of hot star winds,
coronal (solar-type) winds, and the cool winds of evolved, 
late-type stars (Figure~\ref{fig:hr_diag}).

\subsection{Hot Star Winds}

Stars earlier than spectral class B3 typically have high velocity 
winds, and significant mass-loss rates. Radiation pressure is 
substantial in these stars since the (large) opacity of 
iron-group metals peaks in the mid-ultraviolet where the 
radiation field is strong, and so $P_{\rm rad}$ 
is the dominant term in the momentum equation. \citet*{PDB_cas75}
showed that the winds of hot stars are driven by the force of 
radiation on optically thick atomic lines in regions above the 
photosphere, where the continuum is optically thin.

\subsection{Coronal (Solar-type) Winds}

Main sequence stars later than about F0 have substantial convective
zones, which interact with stellar rotation to produce magnetically
active photospheres. Acoustic waves and turbulence generated by 
convective ``noise'' propagate outward into the chromosphere,
where the acoustic waves steepen into shocks and dissipate, producing
localized, transient heating of the chromospheric gas (and the
non-magnetic chromospheric line emission). Gas in the corona, lying
above the chromosphere, is heated to million degree temperatures by
dissipation of magnetic flux and reconnection of magnetic flux 
tubes tied by their photospheric footprint to convective motions. 
Coronal winds are tenuous with a mass-loss rate too small to be of 
evolutionary importance. These are driven mainly by the thermal pressure 
gradient of the hot corona, the \citet{PDB_par58} mechanism, although 
dynamic events in the corona (magnetic reconnection events, flares, and 
coronal mass ejections) also contribute to the solar wind mass flux
(and presumably to that of similar stars also).

\subsection{Cool Winds of Late-Type Stars}

The red supergiant binary $\alpha$~Her was first shown, by 
\citet{PDB_deu56}, to be losing mass via a stellar wind. 
The intervening decades 
of research have failed to produce a viable physical process that
explains the mass-loss process in red giants and 
supergiants \citep{PDB_jud91}. In these stars, 
mass loss occurs via a low velocity 
(typically $\sim$20--50 km s$^{-1}$), massive ($10^{-9}$ to 
$10^{-5}$ M$_\odot$ yr$^{-1}$) wind. This cool wind regime is separated
in the HRD from that of coronal winds by the \citet{PDB_lin79}
dividing line (Figure~\ref{fig:hr_diag}), which
lies near spectral class K1 for red giants and around F0
for the most luminous supergiants.
Noncoronal stars show generally weaker chromospheric emission line 
fluxes and a lower degree of ionization than the coronal stars, and 
have little or no detectable X-ray flux. There is little evidence 
of emission from gas warmer than about 20,000 K, and 
most of the gas must be much cooler than this.

\section{Possible Mass-Loss Mechanisms in Red Giants and Supergiants}

The most stringent constraint on possible mass-loss mechanisms is
provided by the slowly accelerating nature of the wind. Consequently,
most of the work done in driving these cool star winds goes into
lifting material out of the stellar potential and not into
the kinetic energy of the ejecta. This implies an effective driving
force that only slightly exceeds gravity over distances of several
stellar radii.

\subsection{Pulsation}

Pulsation plays a dominant role in initiating mass loss in Mira
variables, OH/IR stars and other cool, evolved stars with optically
thick, dusty winds \citep{PDB_bow88}. These are mostly highly 
evolved asymptotic giant branch (AGB) stars, but probably include 
some massive red supergiant (RSG) stars also, such as WOH G64.
\Citet{PDB_loo08} showed that the ratio of wind kinetic energy to
the power supplied by pulsation 
$(\frac{1}{2}\dot{M}v_\infty^2 / \dot{E}_{\rm puls})$ varies
smoothly with pulsation period, being uniformly large for periods exceeding $600$ d. The RSGs blend in 
indistinguishably from the AGBs in this trend, suggesting that
pulsation may be important in RSGs also. This is consistent with
the suggestion of \citet{PDB_jud91} that long period waves
(with periods longer than the acoustic cutoff period) generated by
photospheric variability may drive mass loss.  On the other hand,
\citet{PDB_jos07} state ``the extrapolation of the theory
of mass loss of AGB stars [to RSGs] seems irrelevant as RSGs have
only irregular, small-amplitude variations and are not pulsating
in a similar manner''. In summary, while it is clear that pulsation 
is a leading contender amongst possible wind-driving mechanisms, 
and it is certainly energetically viable, there still is no
definitive observational evidence that the winds of 
(most) RSGs are actually initiated or driven by pulsation. 
Nevertheless, the observational record is sparse, and the role of 
pulsation in RSGs remains very much an open question.

\subsection{Radiation Pressure on Dust Grains}

Dust grain opacities, unlike atomic opacities, peak in the 
infrared, and so ensure good coupling of the radiation field to 
the circumstellar material. As such, dust-driven winds (usually
initiated by pulsation) have been 
among the most popular of proposed mechanisms explaining
mass loss in red giants and supergiants, e.g. \citet{PDB_laf91},
\citet{PDB_hof08}.
There is much evidence that dust-driven winds are the operative
mechanism in Miras, OH/IR stars, and the optically thick, dusty 
winds of other highly evolved AGB stars (and possibly a few
RSGs such as WOH G64). But these conclusions can not possibly
be generalized to the very different conditions found in the
optically thin winds of RSGs (and first ascent red giant branch 
stars). Some RSGs have relatively little dust, and where the dust
emission has been spatially resolved by IR interferometry
in the case of the M2 supergiant $\alpha$~Ori \citep{PDB_bes96}, 
the observed dust shell lies at a distance of $\sim 30 R_\ast$, 
far exterior to the empirically inferred wind acceleration region 
\citep*{PDB_har01}. Furthermore, VV Cep, another M2 supergiant 
in a binary system and spectral proxy of $\alpha$~Ori, contains 
little or no dust, yet has a comparable mass-loss rate 
to that of $\alpha$~Ori. Similar arguments were advanced years
ago by \citet{PDB_gol79} and \citet{PDB_jud91}, and
the passage of time has not diminished their logic. Radiation
pressure on dust grains can not play a significant role in
driving the optically thin winds of RSGs.

\subsection{Dissipation of Alfv\'{e}n Waves}

One of the most plausible mechanisms proposed to drive
mass loss in RSGs has been the dissipation of Alfv\'{e}n waves.
Unlike acoustic waves, Alfv\'{e}n waves can have large damping
lengths that are far more effective in transferring energy and
momentum to the circumstellar gas over distances of the
order of a stellar radius, as needed to produce stellar winds 
consistent with observations \citep{PDB_har80}. Alfv\'{e}n wave 
dissipation is predicted to result in extended, warm chromospheres
in low gravity stars --- a result that was thought to be consistent
with observation. \citet{PDB_har84} constructed 
an Alfv\'{e}n wave model appropriate for
$\alpha$~Ori which predicts a warm wind with electron temperatures
peaking at $T_{\rm e} \sim 8000$ K at $4 R_\ast$ ($R_\ast$ is the radius 
of the stellar photosphere), and remaining above 5000 K out to 
$10 R_\ast$. However, these results are in strong disagreement with 
the empirical model of \citet{PDB_har01}, in which the wind reaches
a maximum temperature of just 4000 K in the lower chromosphere,
then falls to 2500 K at $3 R_\ast$ and to 800 K by $10 R_\ast$,
with a nearly constant ionization fraction of 
$n_{\rm e}/n_{\rm H} \approx 10^{-3.5}$. It seems unlikely
that Alfv\'{e}n waves, which couple only to
ionized gas, could transfer the necessary energy and momentum 
to the wind in a nearly neutral gas that must be highly damping.

\subsection{Acoustic Waves}

{\sloppy
Photospheric convection in late-type stars generates acoustic
turbulence, and some of this acoustic energy propagates upward into
the chromosphere. \citet{PDB_ath79} showed, from solar UV observations,
that the solar acoustic wave flux is (barely) sufficient to heat the
chromosphere, but not the corona. \citet{PDB_har80} computed
acoustic flux models for low gravity stars and concluded that
``the primary dynamical effect of propagating acoustic waves and
the shocks that develop from them is to extend the essentially
hydrostatic portions of the lower atmosphere, rather than drive a 
steady wind''. \citet{PDB_cun90} confirmed, using a 1-D hydrodynamic code, 
that short period acoustic waves dissipate and heat the chromosphere
but do not contribute significantly to mass loss. 
More sophisticated hydrodyamical modeling of the solar chromosphere 
carried out by \citet{PDB_car95} further confirmed this outcome.
The conclusion: short period acoustic waves heat chromospheres 
but do not cause significant mass loss.}

\subsection{Radiation Pressure on Atomic and Molecular Lines}

The coupling of the cool radiation field of late-type stars
to the line opacity of circumstellar gas is poor, and 
so the resulting radiative force is small. Molecules,
with strong infrared bands, such as carbon monoxide, provide a
better match, but their overall abundance is small, and 
hence their contribution to the radiative force is limited.
Typically, empirical models of the outer atmospheres of RSGs
give a radiative acceleration of the stellar gas of a few times 
$10^{-2}$ cm s$^{-2}$, or $\log a_{\rm rad} \sim -2$. This is
1.5 dex below the $\log g$ of the lowest gravity supergiant.
Radiation pressure on the gas alone can not drive a stellar wind
in red giants or supergiants.

\section{A Tale of Four Stars}

Ultimately, the nature of the driving force accelerating cool stellar
winds will be resolved by observation. To this end, I have analyzed
ultraviolet observations of three bright red supergiants observed with
the {\em HST Goddard High Resolution Spectrograph (GHRS)} and the 
{\em Space Telescope Imaging Spectrograph (STIS)}. With
these results, and including the analysis of \citet{PDB_har01} for
$\alpha$~Ori, we can present comparative empirical results for the
following four RSGs:
\begin{itemize}
\itemsep=-5pt
\item $\lambda$~Vel -- the brightest K~supergiant in the sky: \\
      \rule{1.0in}{0in} $V= 2.21$ mag, \,K4~Ib
\item $\alpha$~Ori -- the brightest M~supergiant in the sky: \\
      \rule{1.0in}{0in} $V= 0.50$ mag, \,M2~Iab
\item $\zeta$~Aur -- the brightest eclipsing K~supergiant binary: \\
      \rule{1.0in}{0in} $V= 3.75$ mag, \,K4~Ib + B5~V, \,$P= 2.66$ yr
\item VV~Cep -- the brightest eclipsing M~supergiant binary: \\
      \rule{1.0in}{0in} $V= 4.91$ mag, \,M2~Iab + B, \,$P= 20.34$ yr
\end{itemize} 
The binaries are of particular interest because, in principle, they
provide the opportunity of obtaining spatially resolved observations
of the outer atmospheres and winds of these red supergiants.

\subsection{\boldmath $\lambda$~Velorum}

The UV spectrum of this red supergiant, observed in 1994 by 
{\em HST/GHRS}, contains numerous emission lines
of singly ionized metals, especially Fe\,{\sc ii}. Optically thin 
lines have a pure emission profile, but optically thicker
lines show self-reversed absorption. This absorption feature 
forms just shortward of the line centre for slightly optically thick 
lines, and deepens and shifts blueward with increasing line optical 
depth. We interpret the velocity at the blueward edge of the 
absorption trough to be the velocity of the wind where the line
becomes optically thin and the underlying emission profile (which
forms close to the star) starts to show through. From this, we directly
infer $v(N_{\rm H})$, where $v$ is the wind velocity and $N_{\rm H}$ is the
hydrogen column density of the wind. Then, assuming a spherical steady
wind with radial hydrogen column density $N_{\rm H}(r)$ and velocity $v(r)$:
\begin{equation}
r = \frac{\dot{M}}{4\pi\mu m_{\rm H} \int_0^{N_{\rm H}(r)} v(N_{\rm H})\, dN_{\rm H}} 
\label{r-eqn}
\end{equation}
and so $r(v)$ is found. Here $\mu$ is the mean particle mass, and $m_{\rm H}$ 
is the hydrogen atom mass. Inverting this result yields the desired 
empirical wind velocity law $v(r)$ with a mass-loss rate 
$\dot{M} = 1.1 \times 10^{-9}$ M$_\odot$ yr$^{-1}$ and a terminal velocity 
of about $30$ km s$^{-1}$ \citetext{Bennett 2010, in preparation}.

\subsection{\boldmath $\alpha$~Orionis}

The line profile analysis technique used for $\lambda$~Vel does not
work well for $\alpha$~Ori because the terminal velocity of the wind 
($\sim$10 km s$^{-1}$) lies within the chromospheric line 
profile, which is broadened by turbulence
of $\sim$20 km s$^{-1}$. However, another approach is available.
Because $\alpha$~Ori has a large angular diameter 
(the largest of any RSG), its wind can be resolved by the
{\em Very Large Array (VLA)} radio telescope. The resolved
radio continuum is a useful diagnostic because the source function
is thermal ($S_\nu = B_\nu$)  and controlled by local conditions, 
which greatly simplifies modelling of the outer atmosphere and wind.
\citet{PDB_har01} used multi-wavelength {\em VLA} radio continuum
fluxes, constrained by additional published radio and mid-IR fluxes,
to derive an empirical model of the wind of $\alpha$~Ori. 
The \citet{PDB_har01} density model,
assuming a spherical, steady flow, gives a mass-loss rate of 
$\dot{M} = 3 \times 10^{-6}$ M$_\odot$ yr$^{-1}$ and a slow wind 
acceleration with a terminal velocity of $\sim$10 km s$^{-1}$.

\subsection{\boldmath Case Studies of Two Supergiant Binaries: 
         $\zeta$~Aur \& VV~Cep}

Observations of RSGs in binary systems provide a unique opportunity
to derive spatially resolved information about their outer atmospheres.
RSG binaries are coeval systems with stars of 
roughly similar mass. The more massive
star has evolved off the main sequence into the RSG region 
of the HRD, while the less massive component remains on the main sequence 
as a (typically) B~star. In the ultraviolet (UV), essentially
all the flux comes from the hot companion. Since the companion
is very much smaller (in physical size, not mass) than the RSG,
the UV line of sight to the companion samples only a small volume
of the supergiant's wind. Along sightlines close to the supergiant's limb,
a ``chromospheric'' absorption spectrum is seen superimposed upon the 
typically featureless B-type continuum.
For the special case of eclipsing binaries, the sightlines during 
ingress and egress probe deep into the supergiant's atmosphere and
provide an unparalleled look at atmospheric structure on very small
spatial scales (about 1--2\% of the supergiant's radius). 

The potential of these RSG binaries (the $\zeta$~Aur stars
with G--K~supergiant primaries, and the VV~Cep stars with M~supergiant
primaries) has been recognized for decades but never fully realized,
largely because these observations were done in the blue-violet 
region of the spectrum \citep{PDB_wil54,PDB_wri70}. In this
spectral region, both stars typically have comparable fluxes, 
and the composite spectrum must be disentangled before
useful analysis can be done. Also, nearly all of the strong
resonance and low-lying atomic transitions, which are most useful 
for mapping the chromosphere, lie in the UV spectrum inaccessible
from the ground. The situation has changed dramatically in recent
decades starting with the launch of {\em IUE} in 1978, and continuing
with the {\em HST/GHRS} and {\em STIS} spectrographs, and the 
recent {\em Far Ultraviolet Spectroscopic Explorer (FUSE)} mission. 
These space missions revolutionized
stellar astronomy by making the UV spectra of stars accessible.

\subsection{\boldmath $\zeta$~Aurigae}

\begin{figure}[!tb]
\plotfiddle{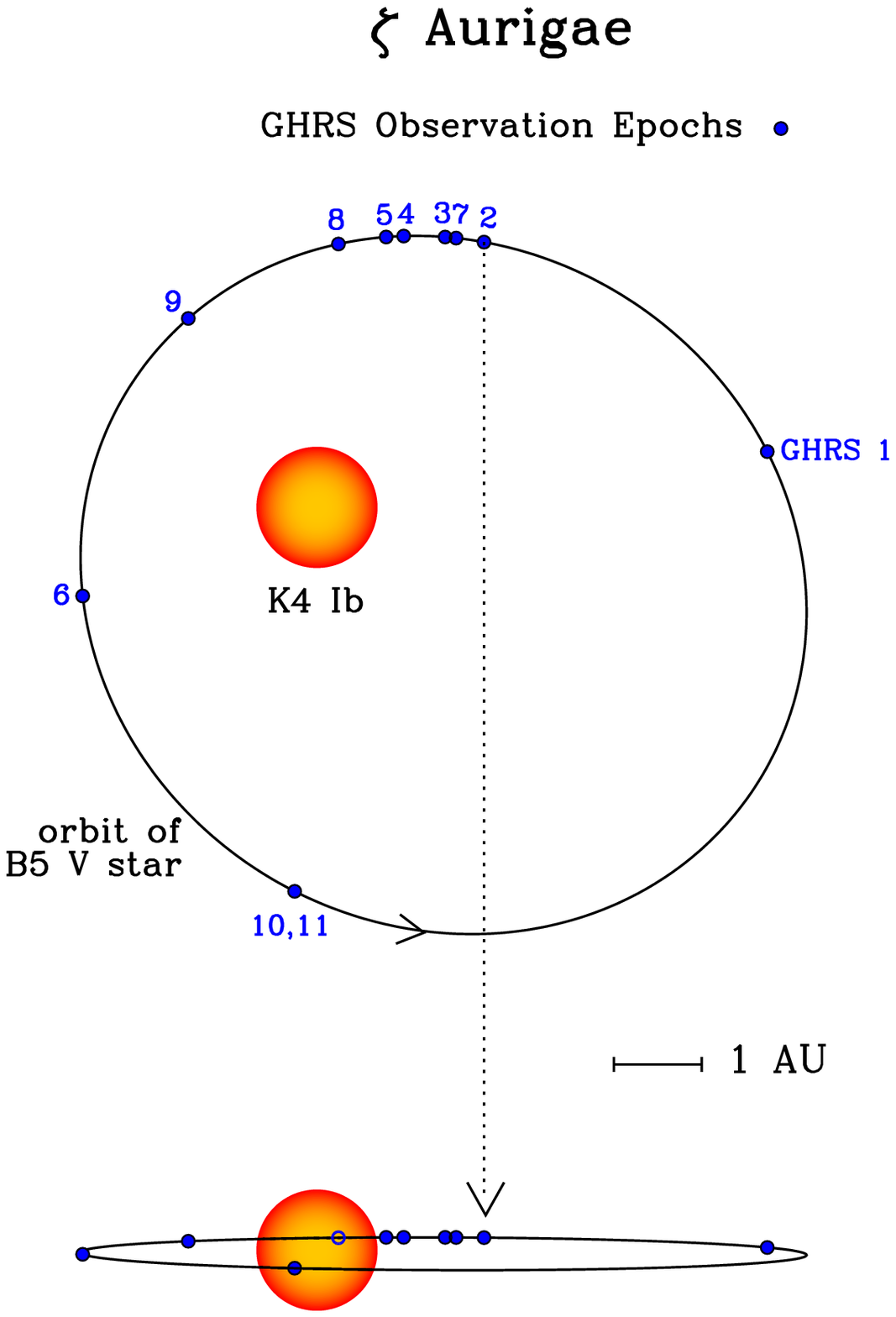}{4.3in}{0}{60}{60}{-242}{-90}
\plotfiddle{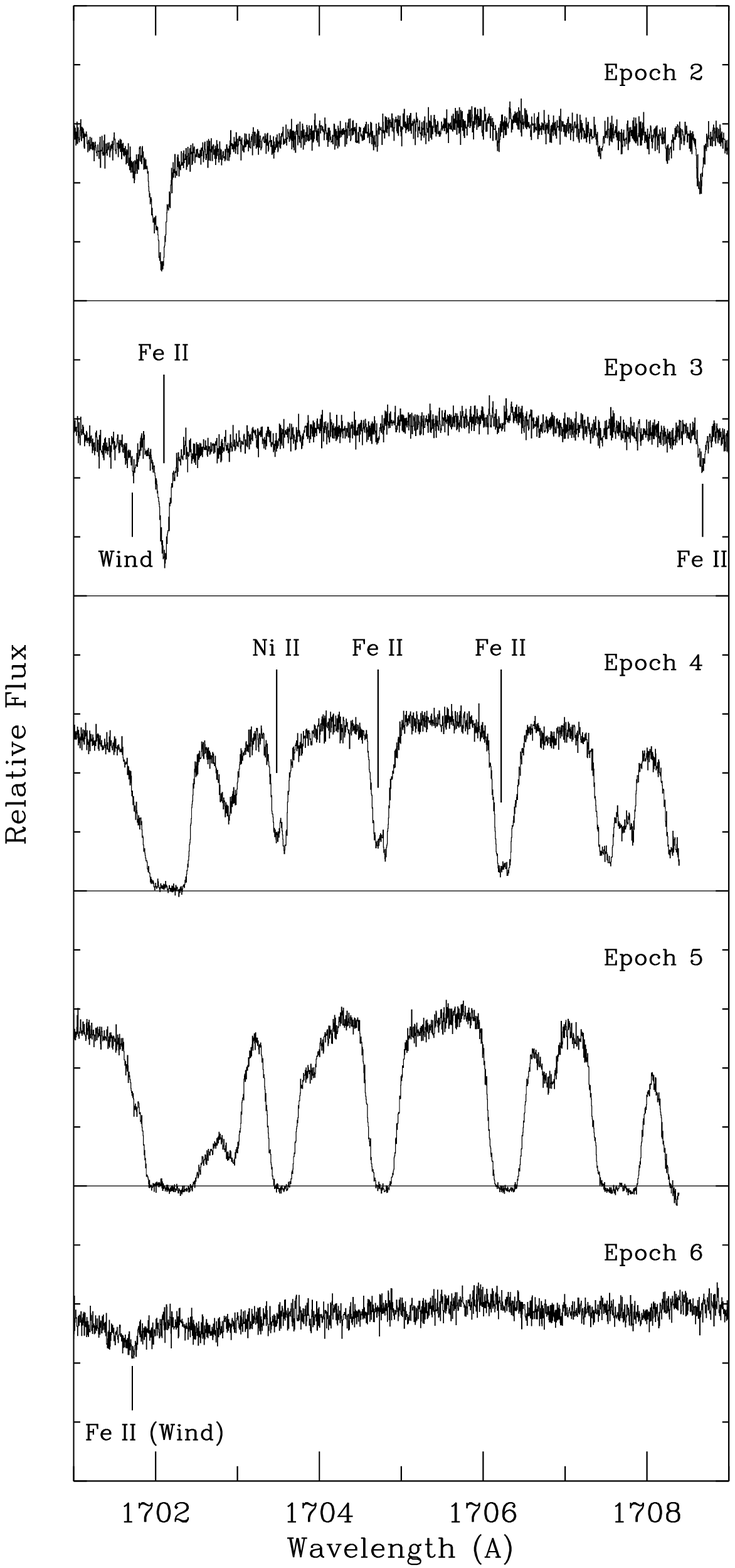}{0.0in}{0}{53}{48}{-7}{-7}
\caption{Left: The relative orbit of $\zeta$~Aurigae, as
viewed in the orbit plane (top), and as projected on the sky (bottom). 
The line of sight to Earth is indicated by the dotted line. 
Positions of the B-type companion at {\em HST} observation epochs 
are shown. Right: The spectrum of $\zeta$~Aur near 1700 \AA\  at various
orbital phases, as observed by the {\em HST/GHRS} echelle spectrograph.
Epochs 4 and 5 show the chromospheric spectrum seen near eclipse.}
\label{fig:orbit}
\end{figure}

In the mid-1990s, $\zeta$~Aur (orbital period $P=2.66$ yr) 
was observed at a series of 11 epochs over 2 orbital periods 
(Figure~\ref{fig:orbit}) by the {\em HST/GHRS}\@. These observations 
permitted the circumstellar density profile to be
mapped, the mean wind law and velocity structure to be derived
(assuming steady, spherical flow), and provided definitive evidence 
of substantial orbit-to-orbit variability in the wind structure. 
This analysis gave a mass loss rate 
$\dot{M}= 5 \times 10^{-9}$ M$_\odot$ yr$^{-1}$
with a terminal velocity of $70$ km s$^{-1}$.
The development of the chromospheric spectrum as eclipse ingress progresses
is also shown in Figure~\ref{fig:orbit}. It is evident that line
profiles show complex structure, especially near eclipse where 
two components are often present in the line core (see Epoch~4 in 
particular). 

\subsection{VV~Cephei}

The UV spectrum of VV~Cep was observed in total eclipse, for the first
time, in November 1997 with the newly commissioned {\em HST/STIS} echelle 
spectrograph. Observations of VV~Cep continued on
for a total of 21 epochs, ending just past first quadrature in 2003.
In eclipse, VV~Cep has a rich emission line spectrum with
more than 2000 features between 1300 and 3150~\AA\ \citep*{PDB_bau07}.
Immediately after egress, VV~Cep shows a complex chromospheric absorption
spectrum, while far from eclipse (near quadrature), an inverse P-Cygni 
``shell'' spectrum becomes prominent. In particular, the chromospheric 
lines sometimes show doubling of the line profiles, 
indicating the presence of complex velocity structure in the 
lower chromosphere. Despite this complexity, reliable line-of-sight
(tangential) column densities through the chromosphere can be obtained
from analysis of the prominent damping wings of the stronger atomic
lines, e.g., Mg\,{\sc ii} $h$ \& $k$, H\,{\sc i} Lyman-$\alpha$, and the stronger
lines of Fe\,{\sc ii}. These tangential column densities can then be used
to construct a density model of the chromosphere and wind of VV~Cep.
The resulting wind model has a mass-loss rate of $\dot{M}$\,=\,1.5\,$\times$\,$10^{-6}$ M$_\odot$ yr$^{-1}$ with a terminal velocity of $v_\infty = 20$ 
km~s$^{-1}$.

\vspace*{-2mm}
\section{The Wind Velocity Law}

We can use these RSG analyses to piece together an idea of
the actual wind velocity law. The binary technique
excels at observations in the chromosphere and the base of the
wind, where pertubations due to the companion are minimal. But, given
the separation of stars is only about $a/R_\ast {\sim6}$ for both
$\zeta$~Aur and VV~Cep (where $a$ is the semi-major axis, $R_\ast$ 
the stellar radius), it is unlikely that the binary
approach provides much useful information at radial distances beyond 
$r\!\sim\!a/2\!\sim\!3 R_\ast$. In this regard the radio continuum flux method 
is complementary since (at present) the spatial resolution is limited to
about $3 R_\ast$ for $\alpha$~Ori.
Similarly, the line profile analysis used for $\lambda$~Vel works
best for the strong lines with well-defined absorption troughs with
well-defined blue edges and so is well-suited for recovering structure 
in the outer wind.

\begin{figure}[htb]
\plotfiddle{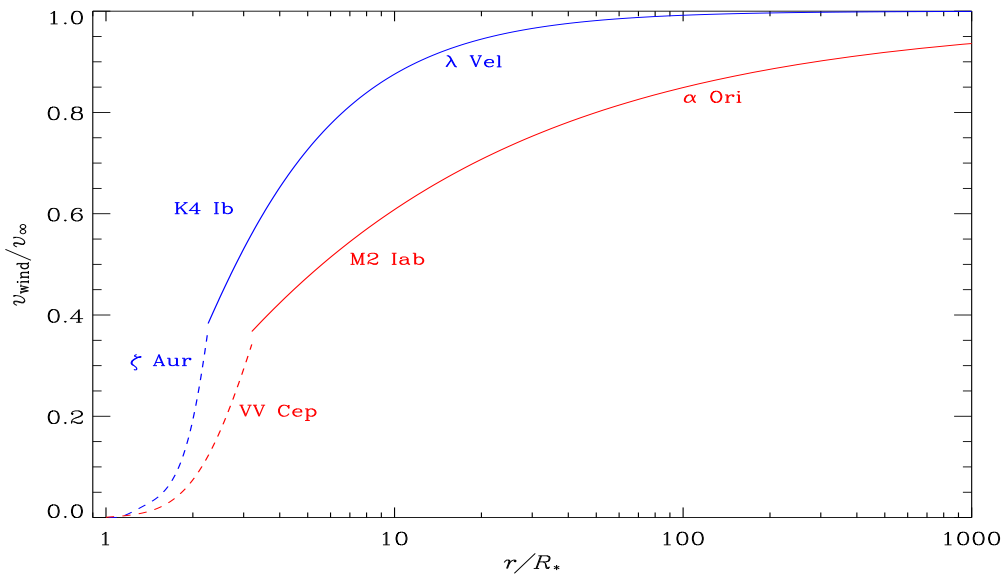}{2.83in}{90}{55}{43}{217}{-4}
\caption{A comparison of wind velocity laws for four RSGs.
The results for single stars (solid) and binaries (dashed) 
are shown together. K~supergiants are shown in blue and M~supergiants
in red. $v_\infty$ is the terminal velocity.}
\label{fig:wind_comp}
\end{figure}

One can get a sense of the complete wind laws for K and
M~supergiants by combining the results of these analyses in a 
single diagram (Figure~\ref{fig:wind_comp}). 
The dashed curves derived for the binary supergiants
are only useful out to a distance of about $3 R_\ast$,
while the solid curves of the single stars are only meaningful
beyond 2--$3 R_\ast$. Presumably combining the useful parts of
each of these curves should approximate the complete velocity law
of K and M~supergiants. 

\vspace*{-2mm}
\section{Line Doubling in the Chromospheric Spectrum}

\begin{figure}[htb]
\plotfiddle{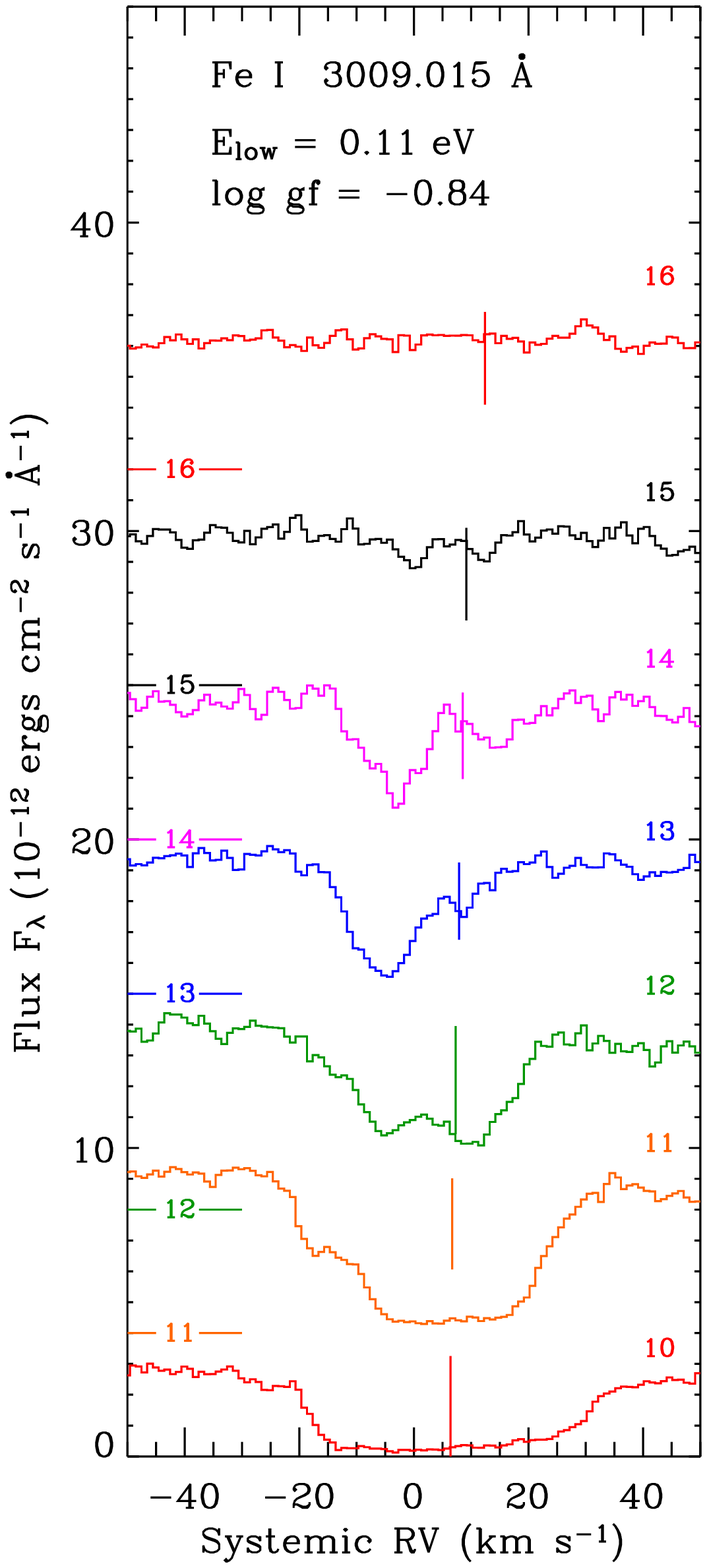}{2.9in}{0}{40}{40}{-194}{-65}
\plotfiddle{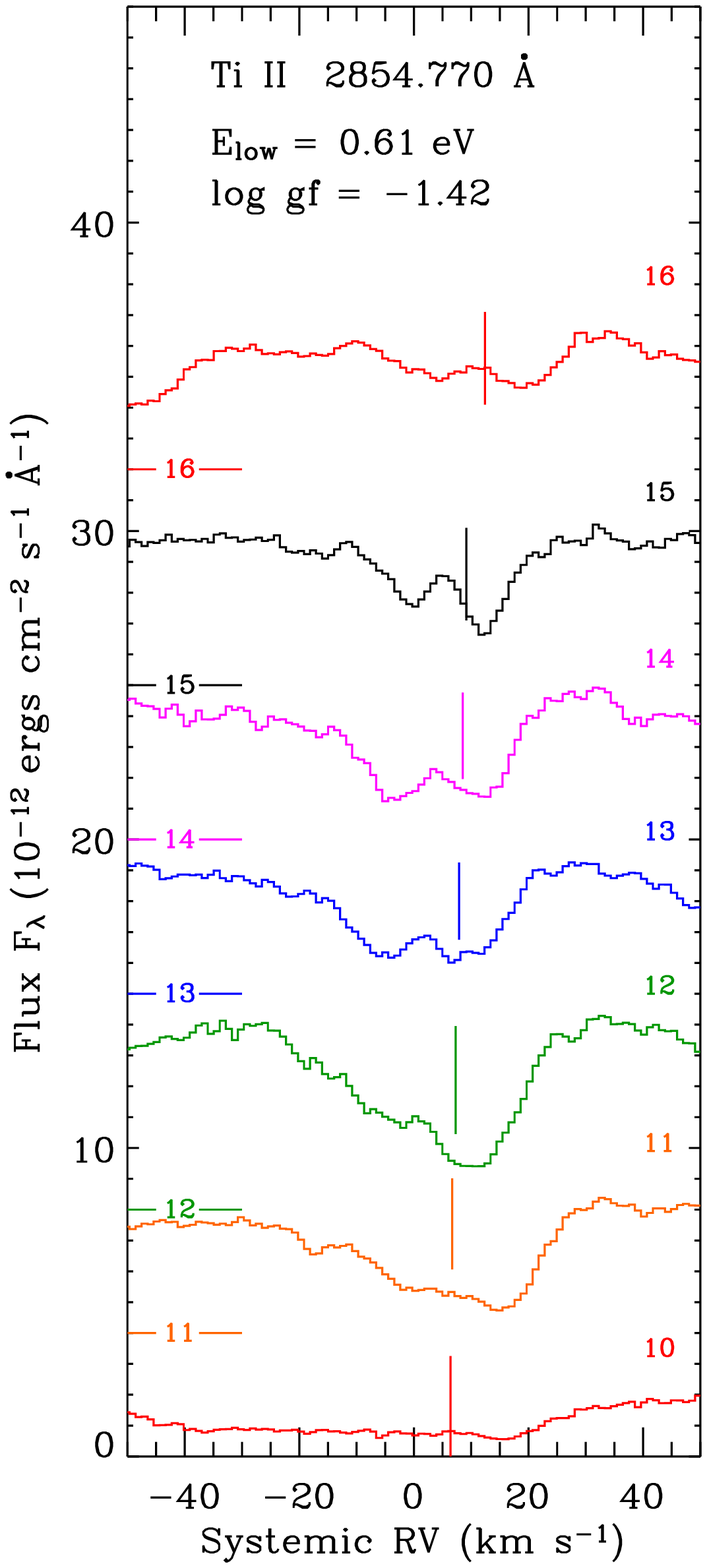}{0.0in}{0}{40}{40}{-62}{-40}
\plotfiddle{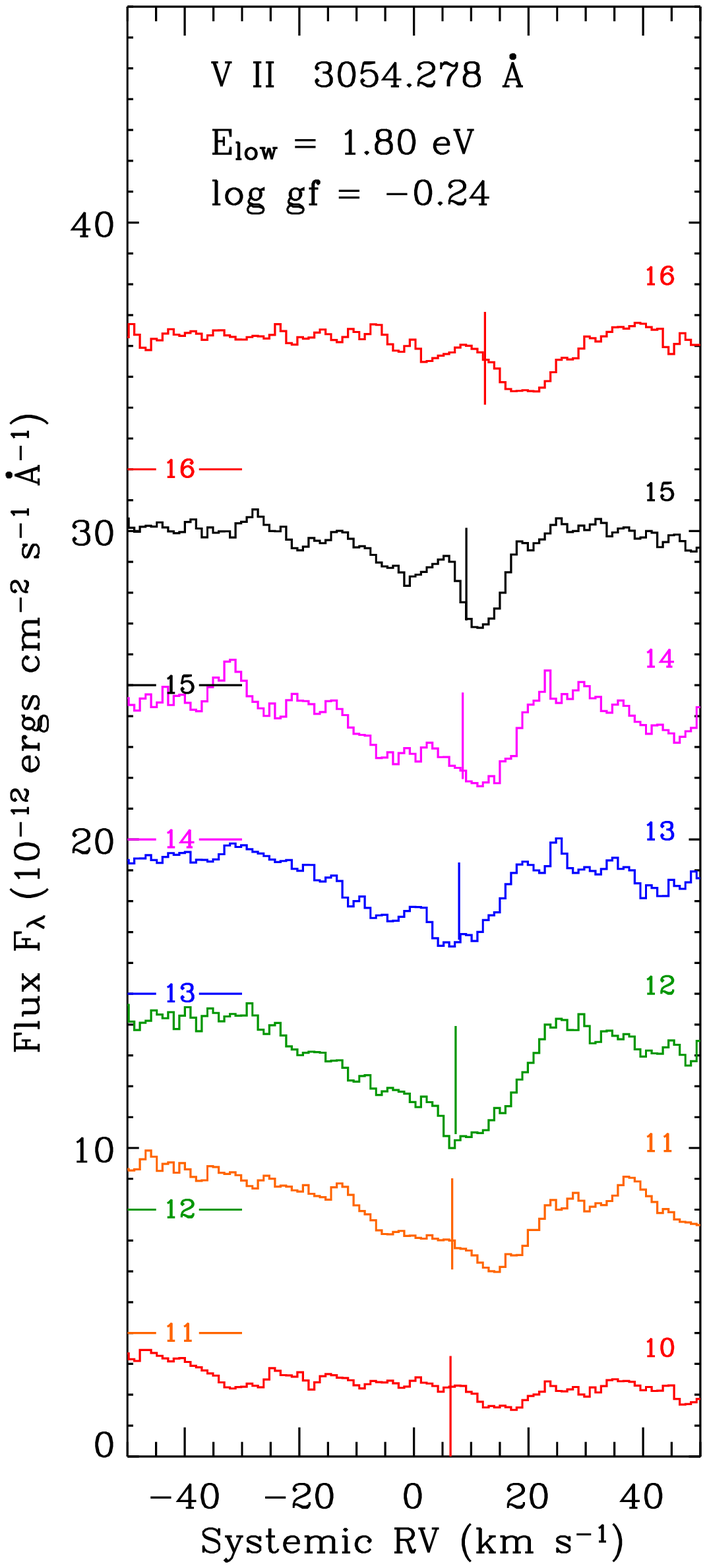}{0.0in}{0}{40}{40}{70}{-15}
\caption{The variation of chromospheric line profiles in VV~Cephei 
with lower-level excitation energy $E_{\rm low}$. The spectra 
are labelled by {\em HST} epoch number on the right. Epoch~10 on 
1998 Dec 1 is the first observation following egress from total eclipse,
while Epoch~16 on 2000 May 18, nearly 1.5 years later, marks the end of 
chromospheric eclipse. Epochs 11--16 are offset vertically, 
with the zero point indicated by the labelled
horizontal line near the left axis. The vertical bars indicate the
M~supergiant's systemic RV at each epoch. Wavelengths of atomic transitions
are in vacuum.}
\label{fig:doubled_lines}
\end{figure}

Despite these results, we still know little about what drives 
these supergiant winds. Some of the assumptions adopted are
manifestly incorrect, e.g., that of a steady wind, given the obvious 
line profile changes in the UV spectrum of $\lambda$~Vel 
between 1990 and 1994 \citep[Fig.~10]{PDB_car99}. The cores of 
chromospheric absorption lines are typically double (e.g., see the
Epoch~4 spectrum, Figure~\ref{fig:orbit}). \citet{PDB_mck59} presented 
the first detailed description of RSG chromospheric line doubling,
as observed in the Ca\,{\sc ii} $K$-line of the RSG 31~Cygni. Line doubling 
is seen in our {\em HST} observations of VV~Cep
(Figure~\ref{fig:doubled_lines}). The 
profile shape and position vary somewhat from epoch to epoch, 
but the separation of the components remains approximately constant. 
For VV~Cep, averaging over several lines and epochs gives 
a mean separation of the components 
of $\Delta v = 14.5 \pm 2.2$ km s$^{-1}$. These components straddle
the supergiant's RV, with the mean blue component at
$v_{\rm blue} = -11.3 \pm 1.8$ km s$^{-1}$ (systemic).

The relative strength of these components
is strongly correlated with lower-level excitation potential. 
Low excitation lines have a dominant blue
component, but the strength of the red component increases with
excitation potential.
Lines from levels above 1 eV show predominantly red components.
The line-doubling phenomenon appears to be 
ubiquitous in RSG chromospheres and needs to be understood before 
more realistic models can be constructed.

\sloppypar

{\acknowledgements \sloppy This research was based on observations with the
NASA/ESA {\em Hubble Space Telescope} obtained at the Space Telescope
Science Institute (STScI). Support for Programs GO-7269, GO-8257, GO-8779 
and GO-9231 (PDB) was provided by NASA through grants from STScI, 
which is operated by AURA, Inc.\ under NASA contract NAS5-26555.}


\end{document}